%% file: main.tex
\pdfoutput=1

\documentclass[12pt,a4paper]{article}
\usepackage{ifthen}
\usepackage{amssymb}
\usepackage{amsfonts}
\usepackage{upgreek}
\usepackage{xspace}
\usepackage{color}
\usepackage{amsmath}
\usepackage{graphicx}
\usepackage{bm}
\usepackage{ctable}
\usepackage{placeins}
\usepackage{grffile}
\usepackage[font=normalsize,subrefformat=parens,labelformat=parens]{subfig}
\usepackage{mathrsfs}
\usepackage{multirow}
\usepackage{relsize}
\usepackage{afterpage}
\usepackage{lmodern}
\usepackage{mciteplus}
\usepackage[bookmarks=true,colorlinks=true,linkcolor=black,citecolor=black,
urlcolor=black]{hyperref}
\usepackage[all]{hypcap}

\newcommand{\xfig}{1}
\ifthenelse{\equal{\xfig}{0}}
{\usepackage[clean,pdf,eps,svgpath=Figures/SVG/,path=Figures/EXP/]{svg}}
{\usepackage[svgpath=Figures/SVG/]{svg}}

\newcommand{\fig}[1]  {\mbox{Fig.\ #1\xspace}}

\renewcommand{\sec}[1]{\mbox{Sect.\ #1\xspace}}

\newcommand{\equ}[1]  {\mbox{Eq.\ #1\xspace}}

\newcommand{\rfr}[1]  {\mbox{Ref.\ #1\xspace}}

\newcommand{\tab}[1]  {\mbox{Table\ #1\xspace}}

\newcommand{\tl}{\ensuremath{\tau} lepton\xspace}
\newcommand{\tls}{\ensuremath{\tau} leptons\xspace}
\newcommand{\thl}{\ensuremath{\tau}-lepton\xspace}

\newcommand{\lep}{\ensuremath{\ell}\xspace}
\newcommand{\had}{\ensuremath{h}\xspace}
\newcommand{\z}{\ensuremath{Z}\xspace}
\newcommand{\w}{\ensuremath{W}\xspace}
\newcommand{\ww}{\ensuremath{{WW}}\xspace}
\newcommand{\ttbar}{\ensuremath{{t\bar{t}}}\xspace}

\newcommand{\jpsi}{\ensuremath{J/\psi}\xspace}
\newcommand{\die}{\ensuremath{ee}\xspace}
\newcommand{\dimu}{\ensuremath{\mu\mu}\xspace}
\newcommand{\dilep}{\ensuremath{\lep\lep}\xspace}
\newcommand{\ditau}{\ensuremath{\tau\tau}\xspace}
\newcommand{\mumu}{\ensuremath{\tau_\mu\tau_\mu}\xspace}
\newcommand{\mue}{\ensuremath{\tau_\mu\tau_e}\xspace}
\newcommand{\emu}{\ensuremath{\tau_e\tau_\mu}\xspace}
\newcommand{\muh}{\ensuremath{\tau_\mu\tau_\had}\xspace}
\newcommand{\eh}{\ensuremath{\tau_e\tau_\had}\xspace}

\newcommand{\pb}{\ensuremath{\mathrm{pb}}\xspace}
\newcommand{\ipb}{\ensuremath{\mathrm{pb}^{-1}}\xspace}

\newcommand{\ifb}{\ensuremath{\mathrm{fb}^{-1}}\xspace}
\newcommand{\tev}{\ensuremath{\mathrm{Te\kern -0.1em V}}\xspace}
\newcommand{\tevc}{\ensuremath{\mathrm{Te\kern -0.1em V\!/}c}\xspace}
\newcommand{\tevcc}{\ensuremath{\mathrm{Te\kern -0.1em V\!/}c^2}\xspace}
\newcommand{\gev}{\ensuremath{\mathrm{Ge\kern -0.1em V}}\xspace}
\newcommand{\gevc}{\ensuremath{\mathrm{Ge\kern -0.1em V\!/}c}\xspace}
\newcommand{\gevcc}{\ensuremath{\mathrm{Ge\kern -0.1em V\!/}c^2}\xspace}
\newcommand{\mev}{\ensuremath{\mathrm{Me\kern -0.1em V}}\xspace}
\newcommand{\mevc}{\ensuremath{\mathrm{Me\kern -0.1em V\!/}c}\xspace}
\newcommand{\mevcc}{\ensuremath{\mathrm{Me\kern -0.1em V\!/}c^2}\xspace}

\newcommand{\lhcb}{LHCb\xspace}
\newcommand{\atlas}{ATLAS\xspace}
\newcommand{\cms}{CMS\xspace}
\newcommand{\pythia}{{\sc Pythia}\xspace}
\newcommand{\powheg}{{\sc Powheg}\xspace}
\newcommand{\herwig}{{\sc Herwig++}\xspace}
\newcommand{\dynnlo}{{\sc Dynnlo}\xspace}
\newcommand{\mstw}{{MSTW$08$}\xspace}
\newcommand{\cteq}{{CTEQ$6$L$1$}\xspace}
\newcommand{\cteqp}{{CTEQ$5$L}\xspace}
\newcommand{\geant}{{\sc Geant4}\xspace}

\newcommand{\ttt}{\ensuremath{\mathrm{TT}}\xspace}
\newcommand{\itt}{\ensuremath{\mathrm{IT}}\xspace}
\newcommand{\ott}{\ensuremath{\mathrm{OT}}\xspace}
\newcommand{\velo}{\ensuremath{\mathrm{VELO}}\xspace}
\newcommand{\ecal}{\ensuremath{\mathrm{ECAL}}\xspace}
\newcommand{\hcal}{\ensuremath{\mathrm{HCAL}}\xspace}
\newcommand{\prs}{\ensuremath{\mathrm{PRS}}\xspace}
\newcommand{\spd}{\ensuremath{\mathrm{SPD}}\xspace}

\newcommand{\pt}{\ensuremath{p_\mathrm{T}}\xspace}
\newcommand{\prc}[1]{\ensuremath{(#1)\%}\xspace}
\newcommand{\abs}[1]{\ensuremath{\left|#1\right|}\xspace}
\newcommand{\ssc}{\ensuremath{\mathrm{SS}}\xspace}

\newcommand{\eff}[1]{\ensuremath{{\varepsilon_{#1}}}\xspace}
\newcommand{\iso}{\ensuremath{{I_{p_\mathrm{T}}}}\xspace}
\newcommand{\dphi}{\ensuremath{{|\Delta \Phi|}}\xspace}
\newcommand{\ips}{\ensuremath{{\mathrm{IPS}}}\xspace}
\newcommand{\apt}{\ensuremath{{A_{p_\mathrm{T}}}}\xspace}
\newcommand{\kin}{\ensuremath{\mathrm{kin}}\xspace}
\newcommand{\gec}{\ensuremath{\mathrm{GEC}}\xspace}
\newcommand{\rec}{\ensuremath{\mathrm{rec}}\xspace}
\newcommand{\sel}{\ensuremath{\mathrm{sel}}\xspace}
\newcommand{\bkg}{\ensuremath{\mathrm{bkg}}\xspace}
\newcommand{\trg}{\ensuremath{\mathrm{trg}}\xspace}
\newcommand{\trk}{\ensuremath{\mathrm{trk}}\xspace}
\newcommand{\id}{\ensuremath{\mathrm{id}}\xspace}
\newcommand{\ewk}{\ensuremath{\mathrm{EWK}}\xspace}
\newcommand{\qcd}{\ensuremath{\mathrm{QCD}}\xspace}

\newcommand{\br}{\ensuremath{\mathcal{B}}\xspace}
\newcommand{\lum}{\ensuremath{\mathscr{L}}\xspace}
\newcommand{\acc}{\ensuremath{\mathcal{A}}\xspace}

\begin{document} 

\input{Latex/title}

\input{Latex/intro}

\input{Latex/analysis}

\input{Latex/measurement}

\input{Latex/results}

\input{Latex/conclusion}

\input{Latex/acknowledgements}

\input{Latex/bibliography}

\end{document}

%% file: Latex/title.tex
\newcommand{\thetitle}{A study of the {\boldmath \z} production
  cross-section in {\boldmath $pp$} collisions at {\boldmath $\sqrt{s}
    = 7~\tev$} using tau final states}

\begin{titlepage}
  
  \pagenumbering{roman}
  \vspace*{-1.5cm}
  \centerline{\large EUROPEAN ORGANIZATION FOR NUCLEAR RESEARCH (CERN)}
  \vspace*{1.5cm}
  \hspace*{-0.5cm}
  \begin{tabular*}{\linewidth}{lc@{\extracolsep{\fill}}r}
    \vspace*{-2.7cm}
    \mbox{\!\!\!\includegraphics[width=.14\textwidth]{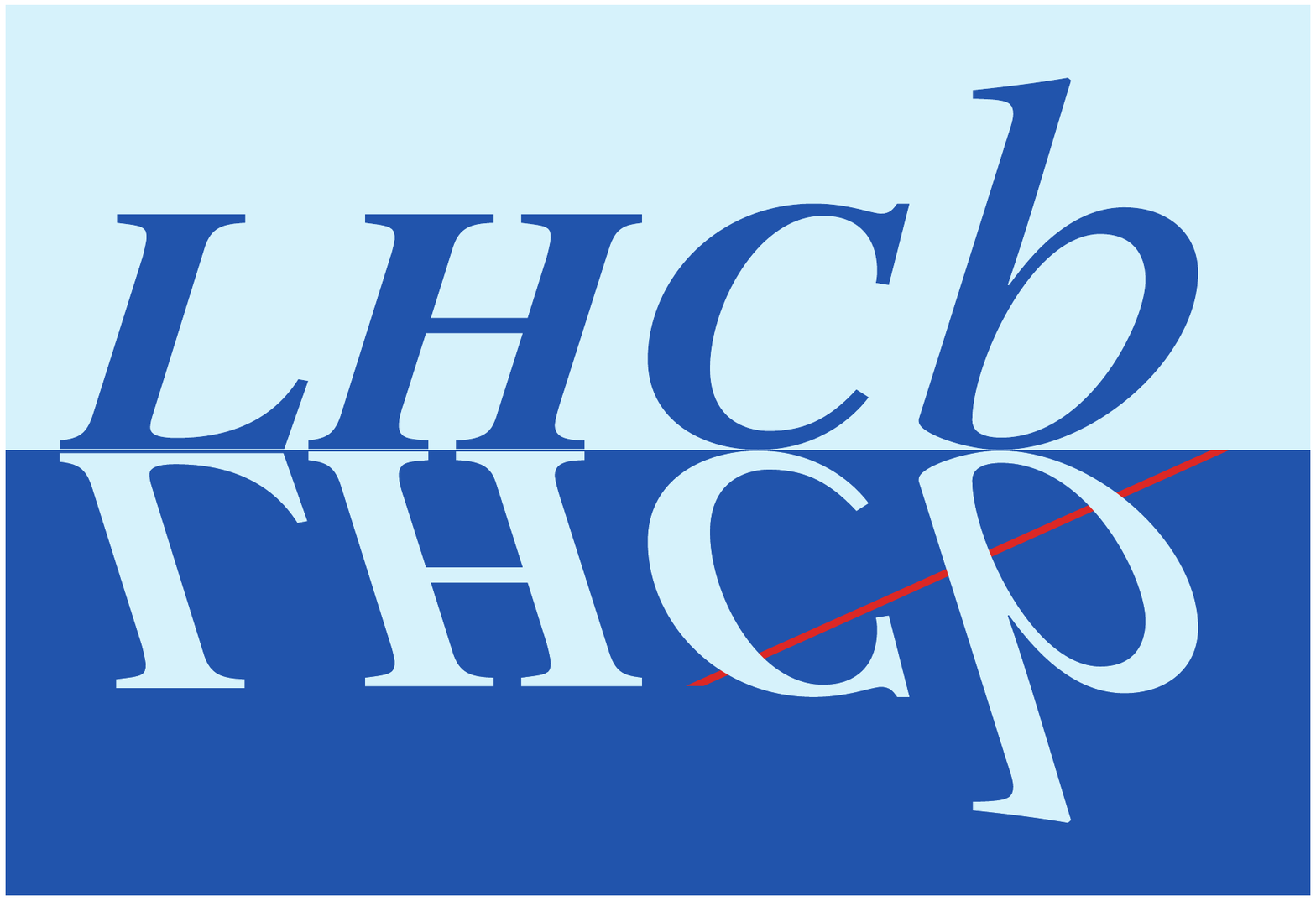}}
    & & \\
    & & CERN-PH-EP-2012-314 \\
    & & LHCb-PAPER-2012-029 \\
    & & October $23$, $2012$ \\
    & & \\
  \end{tabular*}
  
  \vspace*{3.0cm}
  {\bf\huge
    \begin{center}
      \thetitle
    \end{center}
  }
  
  \vspace*{0.5cm}
  \begin{center}
    The \lhcb collaboration\footnote{Authors are listed on the
      following pages.}
  \end{center}
  
  \vspace*{0.5cm}
  \begin{abstract}
    \input{Latex/abstract}
  \end{abstract}
  
  \vspace*{0.5cm}
  \begin{center}
    Submitted to JHEP.
  \end{center}
  \vspace{\fill}
  
\end{titlepage}

\newpage
\setcounter{page}{2}
\mbox{~}
\newpage

\input{Latex/authors}
\cleardoublepage

\setcounter{page}{1}
\pagenumbering{arabic}

%% file: Latex/abstract.tex
\noindent
A measurement of the inclusive $\z \to \ditau$ cross-section in $pp$
collisions at ${\sqrt{s}=7~\tev}$ is presented based on a dataset of
$1.0~\ifb$ collected by the \lhcb detector. Candidates for $\z \to
\ditau$ decays are identified through reconstructed final states with
two muons, a muon and an electron, a muon and a hadron, or an electron
and a hadron. The production cross-section for \z bosons, with
invariant mass between $60$ and $120~\gevcc$, which decay to \tls with
transverse momenta greater than $20~\gevc$ and pseudorapidities
between $2.0$ and $4.5$, is measured to be $\sigma_{pp \to \z \to
  \ditau} = 71.4 \pm 3.5 \pm 2.8 \pm 2.5~\pb$; the first uncertainty
is statistical, the second is systematic, and the third is due to the
uncertainty on the integrated luminosity. The ratio of the
cross-sections for $\z \to \ditau$ to $\z \to \dimu$ is determined to
be $0.93 \pm 0.09$, where the uncertainty is the combination of
statistical, systematic, and luminosity uncertainties of the two
measurements.

%% file: Latex/authors.tex
\centerline{\large\bf LHCb collaboration}
\begin{flushleft}
\small
R.~Aaij$^{38}$, 
C.~Abellan~Beteta$^{33,n}$, 
A.~Adametz$^{11}$, 
B.~Adeva$^{34}$, 
M.~Adinolfi$^{43}$, 
C.~Adrover$^{6}$, 
A.~Affolder$^{49}$, 
Z.~Ajaltouni$^{5}$, 
J.~Albrecht$^{35}$, 
F.~Alessio$^{35}$
% ...

{\footnotesize \it
% ...
$ ^{5}$Clermont Universit\'{e}, Universit\'{e} Blaise Pascal, CNRS/IN2P3, LPC, Clermont-Ferrand, France\\
$ ^{6}$CPPM, Aix-Marseille Universit\'{e}, CNRS/IN2P3, Marseille, France\\
% ...
$ ^{11}$Physikalisches Institut, Ruprecht-Karls-Universit\"{a}t Heidelberg, Heidelberg, Germany\\
% ...
$ ^{33}$Universitat de Barcelona, Barcelona, Spain\\
$ ^{34}$Universidad de Santiago de Compostela, Santiago de Compostela, Spain\\
$ ^{35}$European Organization for Nuclear Research (CERN), Geneva, Switzerland\\
% ...
$ ^{38}$Nikhef National Institute for Subatomic Physics, Amsterdam, The Netherlands\\
% ...
$ ^{43}$H.H. Wills Physics Laboratory, University of Bristol, Bristol, United Kingdom\\
% ...
$ ^{49}$Oliver Lodge Laboratory, University of Liverpool, Liverpool, United Kingdom\\
% ...
$ ^{n}$LIFAELS, La Salle, Universitat Ramon Llull, Barcelona, Spain\\
% ...
}
\end{flushleft}

%% file: Latex/intro.tex
\section{Introduction}\label{sec:int}

The measurement of the production cross-section for \z
bosons\footnote{Here, the \z is used to indicate production from \z
  bosons, photons, and their interference.} in proton-proton ($pp$)
collisions constitutes an important verification of Standard Model
predictions. Since lepton universality in \z decays has been tested to
better than $1\%$ at LEP~\cite{lep}, any deviation observed at the LHC
would be evidence for additional physics effects producing final state
leptons. In particular, \thl pairs can be important signatures for
supersymmetry, extra gauge bosons, or extra
dimensions~\cite{susy,gauge,extra}. The LHCb experiment has previously
measured the cross-section for $\z \to \dimu$~\cite{lhcb} with both
leptons having transverse momentum (\pt) above $20~\gevc$ and an
invariant mass between $60$ and $120~\gevcc$. Here a complementary
measurement in the decay mode $\z \to \ditau$ is presented. This
measurement extends the $\z \to \ditau$ cross-section measurements
from the central pseudorapidity range covered by \atlas
$\left(\abs{\eta} < 2.4\right)$~\cite{atlas} and \cms
$\left(\abs{\eta} < 2.3\right)$~\cite{cms} into the forward region
covered by the \lhcb experiment $(2 < \eta < 4.5)$.

\section{Detector and datasets}\label{sec:det}

The \lhcb detector~\cite{detector} is a single-arm forward
spectrometer designed for the study of particles containing $b$ or $c$
quarks. The detector includes a high precision tracking system
consisting of a silicon-strip vertex detector (\velo) surrounding the
$pp$ interaction region, a large-area silicon-strip detector (\ttt)
located upstream of a dipole magnet with a bending power of about
$4~\mathrm{Tm}$, and three stations of silicon-strip detectors (\itt)
and straw drift tubes (\ott) placed downstream. The combined tracking
system has a momentum resolution $\Delta p/p$ that varies from $0.4\%$
at $5~\gevc$ to $0.6\%$ at $100~\gevc$, and an impact parameter
resolution of $20~\mu\mathrm{m}$ for tracks with high \pt.

Charged hadrons are identified using two ring-imaging Cherenkov
detectors. Photon, electron and hadron candidates are identified by a
calorimeter system consisting of scintillating-pad (\spd) and
pre-shower detectors (\prs), an electromagnetic calorimeter (\ecal)
and a hadronic calorimeter (\hcal). Muons are identified by a system
composed of alternating layers of iron and multiwire proportional
chambers. The trigger consists of a hardware stage, based on
information from the calorimeter and muon systems, followed by a
software stage that applies a full event reconstruction. The hardware
stage imposes a global event requirement (\gec) on the hit
multiplicities of most sub-detectors used in the pattern recognition
algorithms to avoid overloading of the software trigger by high
occupancy events.

This analysis uses data, corresponding to an integrated luminosity of
$1028 \pm 36~\ipb$, taken at a centre-of-mass energy of $7~\tev$. The
absolute luminosity scale was measured periodically throughout the
data taking period using Van der Meer scans~\cite{scan} where the beam
profile is determined by moving the beams transversely across one
another. A beam-gas imaging method was also used where the beam
profile is determined through reconstructing beam-gas interaction
vertices near the beam crossing point~\cite{gas}. Both methods provide
similar results and the integrated luminosity is determined from the
average of the two, with an estimated systematic uncertainty of
$3.5\%$~\cite{lumi}. The primary systematic uncertainty of $2.7\%$ is
due to the beam current measurement, shared between the two methods.

Simulated data samples are used to develop the event selection,
determine efficiencies, and estimate systematic uncertainties. Each
sample was generated using an \lhcb configuration~\cite{tune} of
\pythia $6.4$~\cite{pythia6} with the \cteq leading-order PDF
set~\cite{cteq} and passed through a \geant~\cite{geanta, *geantb}
based simulation of the \lhcb detector~\cite{gauss}. Trigger emulation
and full event reconstruction were performed using the \lhcb
reconstruction software~\cite{brunel}. Additional samples, without
detector simulation or event reconstruction, are used to study the
signal acceptance and were generated using \pythia
$8.1.55$~\cite{pythia8}, \herwig $2.5.1$~\cite{herwig}, and \herwig
with the \powheg method~\cite{powheg}.

%% file: Latex/analysis.tex
\section{Event selection}\label{sec:eve} 

The signatures for $\z \to \ditau$ decays considered in this analysis
are two oppositely-charged tracks, consistent with an electron, muon,
or hadron hypothesis, having large impact parameters with respect to
the primary vertex of the event.

Tracks are reconstructed in the \velo and extrapolated to the
\itt/\ott sub-detectors; any \ttt sub-detector hits consistent with
the track are added and a full track fit is performed. Only tracks
with fit probabilities greater than $0.001$ are considered.

Tracks are extrapolated to the calorimeters and matched with
calorimeter clusters.  Electron candidates are required to have a \prs
energy greater than $0.05~\gev$, a ratio of \ecal energy to candidate
momentum, $E / pc$, greater than $0.1$, and a ratio of \hcal energy to
candidate momentum less than $0.05$. Any electron candidate momentum
is corrected using bremsstrahlung photon recovery~\cite{brem}. Since
the \ecal is designed to register particles from $b$-hadron decays,
calorimeter cells with transverse energy above $10~\gev$ saturate the
electronics, and lead to degradation in the electron energy
resolution.

Hadron candidates are identified by requiring the ratio of \hcal
energy to track momentum to be greater than $0.05$. Due to the limited
\hcal acceptance, the candidate track is required to have a
pseudorapidity of $2.25 \leq \eta \leq 3.75$.

Muon candidates are identified by extrapolating tracks to the muon
system downstream of the calorimeters and matching them with
compatible hits. Muon candidates are required to have a hit in each of
the four stations and consequently will have traversed over $20$
hadronic interaction lengths of material.

The data have been collected using two triggers: a trigger which
selects muon candidates with a \pt greater than $10~\gevc$; and a
trigger which selects electron candidates with \pt greater than
$15~\gevc$.

The analysis is divided into five streams, labelled \mumu, \mue, \emu,
\muh, and \eh, defined such that the streams are exclusive. The first
\tl decay product candidate is required to have $\pt > 20~\gevc$ and
the second is required to have $\pt > 5~\gevc$. The following
additional kinematic and particle identification requirements are
specific to each analysis stream:
\begin{itemize}
\item \mumu requires two oppositely-charged muons where at least one
  triggered the event. The muon with the larger \pt is considered as
  the first \tl decay product candidate.
\item \mue requires a muon that triggered the event and an
  oppositely-charged electron.
\item \emu requires an electron and an oppositely-charged muon with
  $\pt < 20~\gevc$. Either lepton can trigger the event.
\item \muh requires a muon that triggered the event and an
  oppositely-charged hadron.
\item \eh requires an electron that triggered the event and an
  oppositely-charged hadron.
\end{itemize}

In $pp$ collisions the cross-section for hadronic QCD processes is
very large. These events can pass the above requirements either due to
semileptonic $c$- or $b$-hadron decays or through the
misidentification of hadrons as leptons.

Signal decays, coming from an on-shell \z, tend to have back-to-back
isolated tracks in the transverse plane with a higher invariant mass
than tracks in QCD events. The absolute difference in azimuthal angle
of the two \tl decay product candidates, \dphi, is required to be
greater than $2.7~\mathrm{radians}$ and their invariant mass is
required to be greater than $20~\gevcc$.

Tracks in QCD events also tend to be associated with jet activity, in
contrast to signal events where they are isolated. An isolation
variable, \iso, is defined as the transverse component of the
vectorial sum of all track momenta that satisfy $\sqrt{\Delta \phi^2 +
  \Delta \eta^2} < 0.5$, where $\Delta \phi$ and $\Delta \eta$ are the
differences in $\phi$ and $\eta$ between the \tl decay product
candidate and the track. The track of the \tl decay product candidate
is excluded from the sum. Both \tl decay product candidates are
required to have $\iso < 2~\gevc$ for the \mumu, \mue, and \emu
analysis streams and $\iso < 1~\gevc$ for \muh and \eh due to the
larger \qcd backgrounds.

The lifetime of the \tl is used to separate signal from prompt
backgrounds. The signed impact parameter for a track is defined as the
magnitude of the track vector of closest approach to the primary
vertex signed by the $z$-component of the cross product between this
vector and the track momentum. The impact parameter significance,
\ips, is then defined as the absolute sum of the signed impact
parameters of the two \tl decay product candidates, divided by their
combined uncertainty. The \ips is required to be greater than $9$ for
the \mumu, \muh, and \eh analysis streams while no \ips requirement is
placed on the \mue or \emu streams.

In the \mumu analysis stream an additional background component arises
from $\z \to \dimu$ events. This produces two muons with similar \pt,
most of which also have an invariant mass close to the \z mass. In
contrast, signal events tend to have unbalanced \pt and a lower
invariant mass due to unreconstructed energy from neutrinos and
neutral hadrons. The \pt asymmetry, \apt, is defined as the absolute
difference between the \pt of the two candidates divided by their
sum. For the \mumu analysis stream the \apt is required to be greater
than $0.3$ and the di-muon invariant mass must lie outside the range
$80 < M_{\mu\mu} < 100~\gevcc$.

\section{Background estimation}\label{sec:bac}

\begin{figure}
  \newlength{\plotwidth}
  \newlength{\plotsep}
  \newlength{\capwidth}
  \newlength{\capsep}
  \setlength{\plotwidth}{0.5\textwidth}
  \setlength{\plotsep}{-0.1cm}
  \setlength{\capwidth}{2.4cm}
  \setlength{\capsep}{-1.6cm}
  \begin{center}
    \vspace{-1.9cm}
    \centerline{
      \begin{tabular}{@{\extracolsep{-0.6cm}}cc}
        \subfloat[]{\label{fig:mumu}}\hspace{\capwidth} &
        \subfloat[]{\label{fig:mue}}\hspace{\capwidth} \\[\capsep]
        \includesvg[width=\plotwidth,name=Fig.\arabic{svgfigure}a]
        {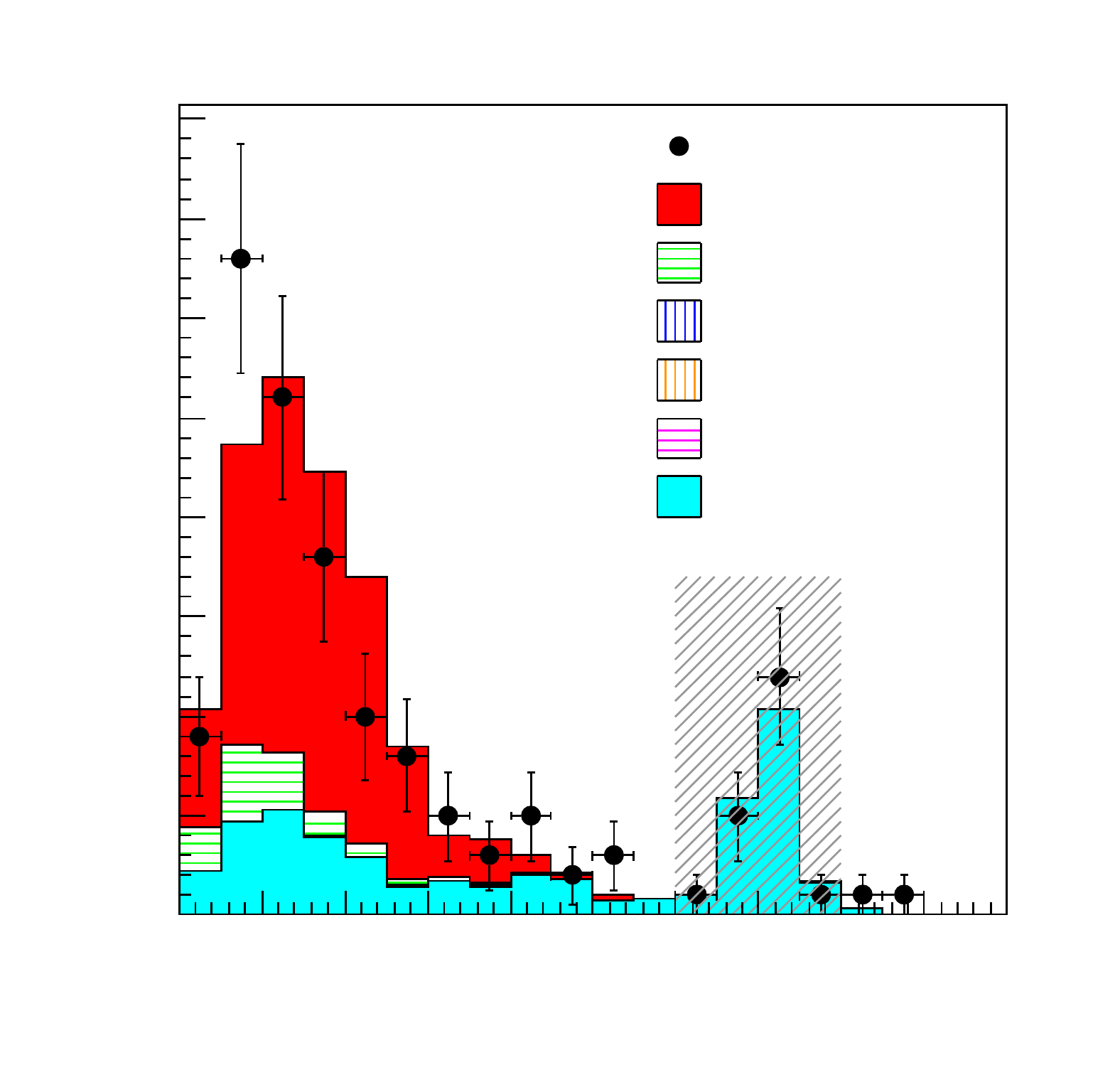} &
        \includesvg[width=\plotwidth]{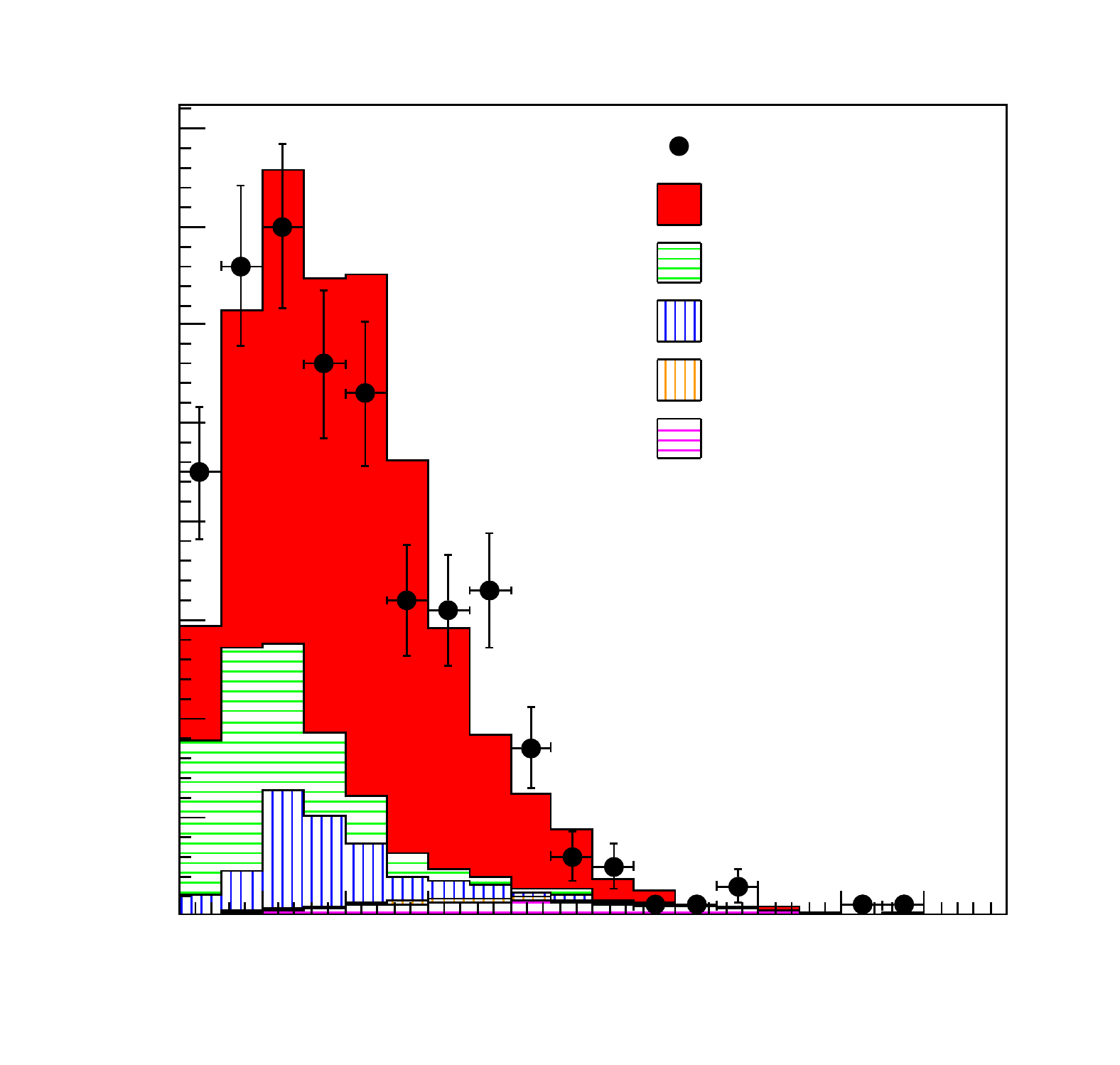}
        \\[\plotsep]
        \subfloat[]{\label{fig:emu}}\hspace{\capwidth} &
        \subfloat[]{\label{fig:muh}}\hspace{\capwidth} \\[\capsep]
        \includesvg[width=\plotwidth,name=Fig.\arabic{svgfigure}c]
        {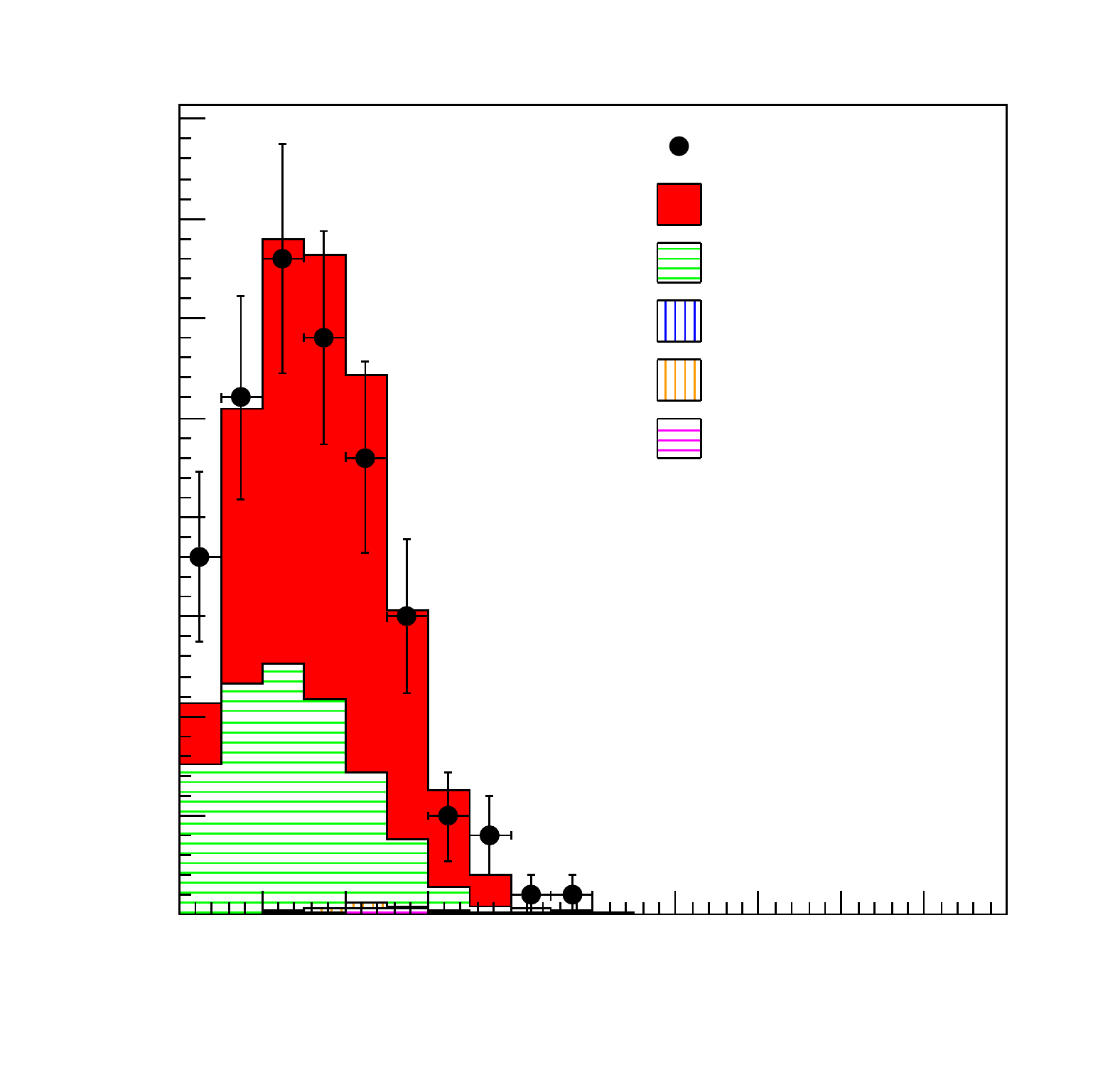} &
        \includesvg[width=\plotwidth]{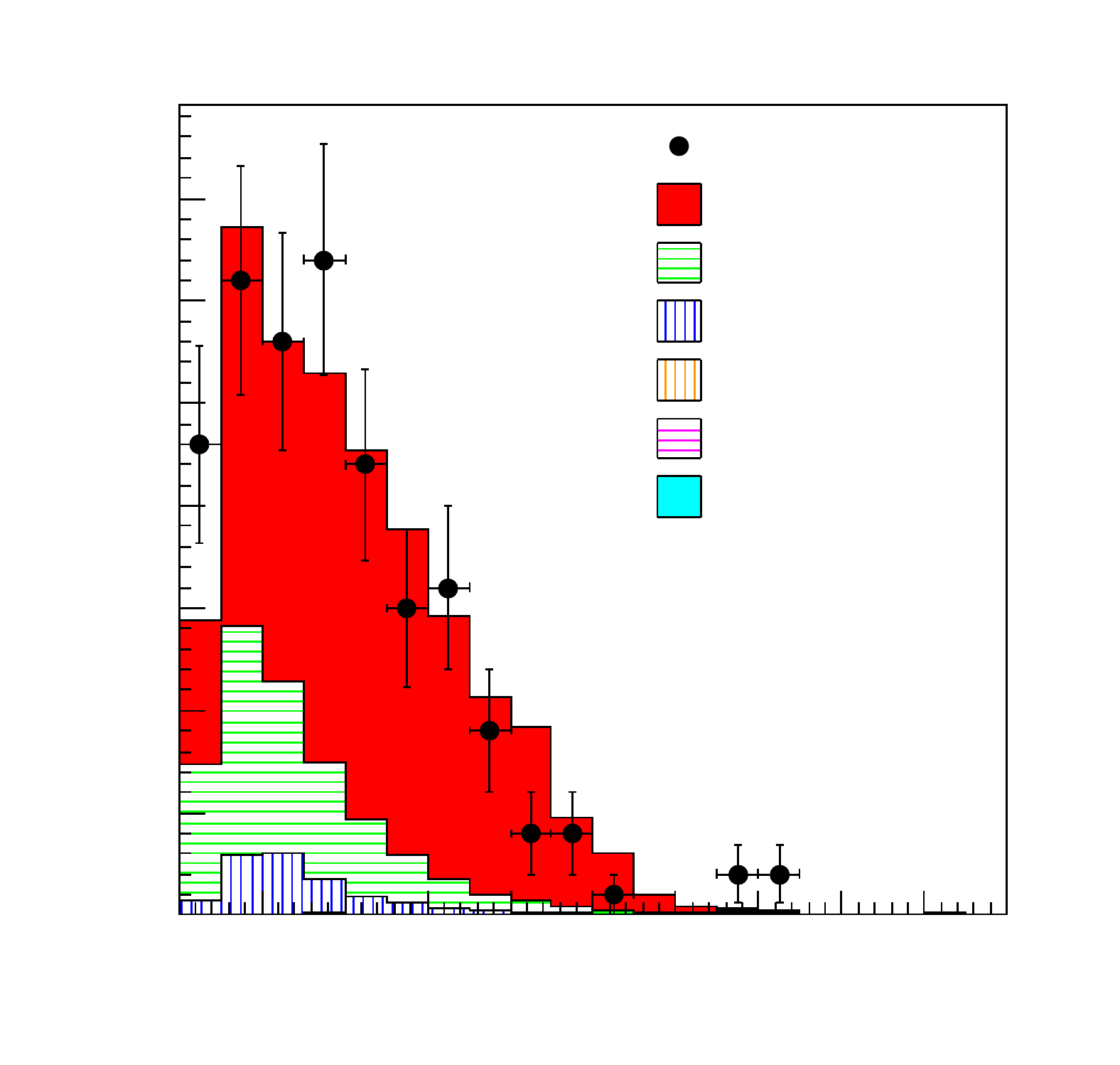}
        \\[\plotsep]
        \subfloat[]{\label{fig:eh}}\hspace{\capwidth} & \\[\capsep]
        \includesvg[width=\plotwidth]{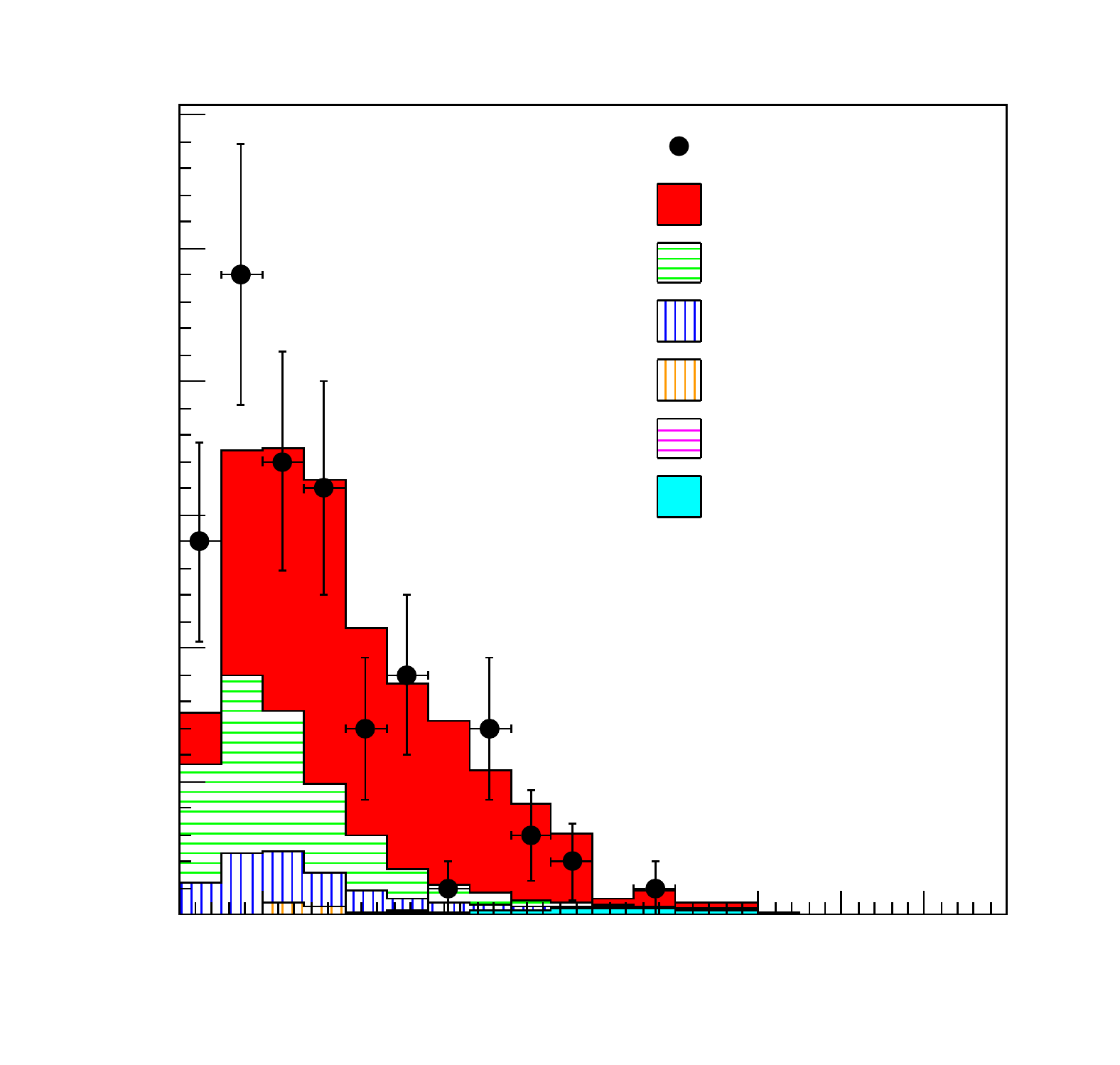} & \\
      \end{tabular}}
  \end{center}
  \vspace{-0.95cm}
  \caption{Invariant mass distributions for the
    \protect\subref{fig:mumu} \mumu, \protect\subref{fig:mue} \mue,
    \protect\subref{fig:emu} \emu, \protect\subref{fig:muh} \muh, and
    \protect\subref{fig:eh} \eh candidates with the excluded mass
    range indicated for \mumu. The $\z \to \ditau$ simulation (solid
    red) is normalised to the number of signal events. The \qcd
    (horizontal green), electroweak (vertical blue), and \z (solid
    cyan) backgrounds are estimated from data. The \ttbar (vertical
    orange) and \ww (horizontal magenta) backgrounds are estimated
    from simulation and generally not visible.\label{fig:mas}}
\end{figure}

The invariant mass distributions for the selected $\z \to \ditau$
candidates, the simulated signal, and the estimated backgrounds for
the five analysis streams are shown in \fig{\ref{fig:mas}}, where no
candidates are observed with a mass above $120~\gevcc$. Five types of
background have been considered: generic QCD; electroweak, where a
high \pt lepton is produced by a \w or \z boson and the second
candidate \tl decay product is misidentified from the underlying
event; \ttbar, where two hard leptons are produced from top decays;
\ww, where each \w decays to a lepton; and $\z \to \dilep$ for the
\mumu, \muh, and \eh streams, where for the \muh stream a single muon
is misidentified as a hadron and for the \eh stream an electron is
misidentified. The \ttbar and \ww backgrounds are estimated using
simulation and found to be small for all final states.

The QCD and electroweak backgrounds are estimated from data. A
signal-depleted control sample is created by applying all selection
criteria but requiring that the \tl decay product candidates have the
same-sign (\ssc) charge. The QCD and electroweak background events in
this sample, $N_\qcd^\ssc$ and $N_\ewk^\ssc$, are obtained by fitting
template shapes to the distribution of the difference between the \pt
of the first and second \tl decay product candidates. The template
shape for the electroweak background is taken from simulation. To
determine the shape of the QCD contribution, the isolation requirement
is reversed such that $\iso > 10~\gevc$. The number of candidates for
each background category in the signal sample is then calculated as
$N_\qcd = f_\qcd N^\ssc_\qcd$ and $N_\ewk = f_\ewk N^\ssc_\ewk$, where
$f_\qcd$ and $f_\ewk$ are the ratio of opposite-sign to same-sign
events for QCD and electroweak events respectively. Both $f_\qcd$ and
$f_\ewk$ are determined as the ratio of opposite-sign to same-sign
events satisfying the template requirements. The uncertainties on the
\qcd and electroweak backgrounds are estimated by combining the
statistical uncertainty on the fraction with the uncertainties from
the fit used to determine $N_\qcd^\ssc$ and $N_\ewk^\ssc$.

The number of $\z \to \dimu$ background events for the \mumu stream is
obtained by applying all selection criteria except for the $80 <
M_{\mu\mu} < 100~\gevcc$ requirement. This produces a sample with a
clear peak around the \z mass as shown in \fig{\subref*{fig:mumu}}. A
template for $\z \to \dimu$ events is obtained from data by applying
the event selection, but requiring prompt events with $\ips < 1$. The
template is normalised to the number of events within the \mumu sample
with $\ips > 9$ and within the invariant mass range $80 < M_{\mu\mu} <
100~\gevcc$. The $\z \to \dimu$ background is the number of events in
the normalised template outside this mass range. The uncertainty on
this background is estimated from the statistical uncertainty on the
normalisation factor.

The $\z \to \dimu$ process also contributes a small background to the
\muh stream when one of the muons is misidentified as a hadron. This
is evaluated by applying the \muh selection but requiring a second
identified muon rather than a hadron, and scaling this by the
probability for a muon to be misidentified as a hadron. The latter is
found from a sample of $\z \to \dimu$ events that have been selected
by requiring a single well defined muon and a second isolated track,
which give an invariant mass between $80$ and $100~\gevcc$; \prc{0.06
  \pm 0.01} of these tracks pass the hadron identification
requirement.

Similarly, a small $\z \to \die$ background can contribute to the \eh
stream when one of the electrons is misidentified as a hadron. This
is evaluated by applying the \eh selection but requiring a second
identified electron rather than a hadron, and scaling this by the
probability for an electron to be misidentified as a hadron. The
electron mis-identification is found from simulated $\z \to \die$
events to be \prc{0.63 \pm 0.02}.

%% file: Latex/measurement.tex
\section{Cross-section measurement}\label{sec:cro}

The $pp \to \z \to \ditau$ cross-section is calculated within the
kinematic region $60 < M_{\tau\tau} < 120~\gevcc$, $2.0 \leq \eta^\tau
\leq 4.5$, and $\pt^\tau > 20~\gevc$ using

\begin{equation}
  \sigma_{pp \to \z \to \ditau} = \frac{\sum_{i = 1}^{N} 1 /
    \varepsilon_\rec^i - \sum_j N^j_\bkg \langle 1 / \varepsilon_\rec \rangle^j}
  {\lum \cdot \acc \cdot \br \cdot \eff{\sel}}
  \label{equ:xs}
\end{equation}
where $N$ is the number of observed candidates and $N_\bkg^j$ is the
estimated background from source $j$. The integrated luminosity is
given by \lum, \acc is an acceptance and final state radiation
correction factor, \br is the branching fraction for the \thl pair to
decay to the final state, and \eff{\sel} is the selection
efficiency. A summary of these values for each final state is given in
\tab{\ref{tab:mea}}. The reconstruction efficiency, \eff{\rec}, is
calculated using simulation or data for each event, assuming that it
is signal, and depends on the momentum and pseudorapidity of the \tl
decay product candidates.  $\langle 1 / \varepsilon_\rec\rangle^j$
indicates the average value of $1/\eff{\rec}$ for background source
$j$.

\begin{table}
  \caption{Acceptance factors, branching fractions, selection
    efficiencies, numbers of background and observed events for each
    $\z \to \ditau$ analysis stream.\label{tab:mea}}
  \begin{center}
    \small
    \begin{tabular}[t]{c|r|r|r|r|r}
      \toprule
      \multicolumn{1}{c|}{Stream}
      & \multicolumn{1}{c|}{\acc} 
      & \multicolumn{1}{c|}{\br $[\%]$}
      & \multicolumn{1}{c|}{\eff{\sel}}
      & \multicolumn{1}{c|}{$N_\bkg$}
      & \multicolumn{1}{c}{$N$} \\
      \midrule
      \mumu & $0.405 \pm 0.006$ & $3.031 \pm 0.014$ & $0.138 \pm 0.006$ 
      & $ 41.6 \pm 8.5$ & $124$ \\
      \mue  & $0.248 \pm 0.004$ & $6.208 \pm 0.020$ & $0.517 \pm 0.012$
      & $129.7 \pm 4.9$ & $421$ \\
      \emu  & $0.152 \pm 0.002$ & $6.208 \pm 0.020$ & $0.344 \pm 0.016$
      & $ 56.6 \pm 3.3$ & $155$ \\
      \muh  & $0.182 \pm 0.002$ & $16.933 \pm 0.056$ & $0.135 \pm 0.004$
      & $ 53.3 \pm 0.8$ & $189$ \\
      \eh   & $0.180 \pm 0.002$ & $17.341 \pm 0.057$ & $0.082 \pm 0.004$
      & $ 36.6 \pm 0.9$ & $101$ \\
      \bottomrule
    \end{tabular}
  \end{center}
\end{table}

The integrated luminosity of the datasets for the \mumu, \mue, and
\muh samples is $1028 \pm 36~\ipb$, while the \emu and \eh final state
datasets have an integrated luminosity of $955 \pm 33~\ipb$.

\subsection{Acceptances and branching fractions}\label{sec:cro:acc}

The acceptance factor, \acc, is used to correct the kinematics of each
analysis stream to the kinematic region $60 < M_{\tau\tau} <
120~\gevcc$, $2.0 \leq \eta^\tau \leq 4.5$, and $\pt^\tau >
20~\gevc$. This region corresponds to the detector fiducial acceptance
and allows a comparison with the \lhcb $\z \to \dimu$
measurement~\cite{lhcb}. The acceptance factor is taken from
simulation and is defined as the number of $\z \to \ditau$ events
where the generated \tl decay products fulfil the kinematic
requirements described in \sec{\ref{sec:eve}}, divided by the number
of $\z \to \ditau$ events where the generated \tls lie within the
kinematic region defined above.

For each final state the acceptance factors are calculated at
leading-order using fully modelled hadronic decay currents and spin
correlated \tl decays with final state radiation in \pythia $8$ and
\herwig, and at next-to-leading-order using the \powheg method
implemented in \herwig. For \pythia $8$ the \cteqp leading-order PDF
set~\cite{cteqp} was used, while for \herwig the \mstw PDF
set~\cite{mstw} was used. The mean of the maximum and minimum values
from the three generators is taken as the acceptance factor and is
given in \tab{\ref{tab:mea}}. The uncertainty is taken as half the
difference between the maximum and minimum values.

The branching fractions are calculated using the world
averages~\cite{pdg} and are given in \tab{\ref{tab:mea}}. The \tl to
single charged-hadron branching fraction is the sum of all \tl decays
containing a single charged hadron. The final states presented in this
analysis account for $44\%$ of all expected $\z \to \ditau$ decays.

\subsection{Selection efficiency}\label{sec:cro:sel}

The event selection efficiency, $\eff{\sel}$, is the product of the
efficiencies described below. Each efficiency is determined from
either data, or simulation which has been calibrated using data. The
resulting $\eff{\sel}$ for each stream is given in
\tab{\ref{tab:mea}}.

The kinematic efficiency, \eff{\kin}, is obtained from simulation and
is the number of events fulfilling the kinematic requirements of
\sec{\ref{sec:eve}} at both the simulated and reconstructed level
divided by the number of events passing the requirements at the
simulated level. The efficiency is consistent with unity for the \mumu
and \muh analysis streams. For streams involving electrons, \eff{\kin}
is significantly lower due to the saturation of the \ecal. This
results in electrons being reconstructed with lower momenta than their
true momenta due to incomplete bremsstrahlung recovery. In the \mue,
\emu, and \eh streams, \eff{\kin} is \prc{99.3 \pm 1.0}, \prc{66.8 \pm
  1.9}, and \prc{67.0 \pm 1.3} respectively. The uncertainties come
from the statistical uncertainty of the $\z \to \ditau$ simulation and
the calibration of the electron momentum scale which has been obtained
by comparing the \pt spectrum of $\z \to \die$ events in data and
simulation~\cite{zee}.

The efficiency of the isolation requirement, \eff{\iso}, for each
analysis stream is taken from $\z \to \ditau$ simulation, and
calibrated to data by multiplying \eff{\iso} by the ratio of the
efficiency obtained in $\z \to \dimu$ data to $\z \to \mumu$
simulation. The systematic uncertainty on \eff{\iso} is estimated as
the difference between the efficiencies obtained from $\z \to \dimu$
simulation and $\z \to \mumu$ simulation.

The efficiency of the impact parameter significance requirement,
\eff{\ips}, is evaluated from $\z \to \ditau$ simulation. A comparison
of the \ips distributions in $\z \to \dimu$ events from data and
simulation show that the impact parameter resolution is underestimated
by \prc{12\pm 1} in simulation, and so the simulated $\z \to \ditau$
events are corrected by this factor. The systematic uncertainty on
\eff{\ips} is determined by re-calculating the efficiency in $\z \to
\ditau$ simulation with the scale factor varied by its uncertainty.

The efficiency of the azimuthal angle separation requirement,
\eff{\dphi}, and \pt asymmetry efficiency requirement, \eff{\apt}, are
evaluated from simulation. The systematic uncertainty on each is taken
as the difference in the evaluation of these efficiencies in $\z \to
\dimu$ data and simulation, combined in quadrature with the
statistical uncertainty from the $\z \to \ditau$ simulation.

\subsection{Reconstruction efficiency}\label{sec:cro:rec}

The reconstruction efficiency, \eff{\rec}, is the product of the \gec,
trigger, and tracking and identification efficiencies for both \tl
decay product candidates. The tracking efficiency is the probability
for reconstructing the track and the identification efficiency is the
probability for the track to be identified by the relevant
sub-detectors. All efficiencies determined from data have been checked
against simulation and found to agree within the percent level.

The \gec efficiency, \eff{\gec}, is a correction for the loss due to
the rejection by the hardware trigger of events with an \spd
multiplicity of greater than $600~\mathrm{hits}$.  For muon triggered
events, the efficiency has been evaluated to be \prc{95.5 \pm 0.1}
from $\z \to \dimu$ data events using a hardware di-muon trigger with
a relaxed \spd requirement of $900~\mathrm{hits}$. For electron
trigger events, the efficiency is estimated to be \prc{95.1 \pm 0.1}
by comparing the hit multiplicities in $\z \to \dimu$ and $\z \to
\die$ events.

The muon and electron trigger efficiencies, \eff{\trg}, are evaluated
in bins of momentum using a tag-and-probe method on $\z \to \dilep$
data events, which have been selected requiring two reconstructed and
identified muon or electron candidates with an invariant mass
consistent with that of the \z. In the events the triggered lepton is
taken as the tag lepton, and the other as the probe lepton. The event
topologies for $\z \to \dilep$ and $\z \to \ditau$ events are nearly
identical except for the momenta of the final state particles and so
the trigger efficiency is calculated only as a function of the probe
momentum below $500~\gevc$. The trigger efficiency is the fraction of
events where the probe has also triggered, and varies as a function of
probe momentum between $75\%$ and $80\%$ for the muon trigger and
between $62\%$ and $75\%$ for the electron trigger. The trigger
efficiency uncertainty for each bin in momenta is taken as the
statistical uncertainty.

The tracking efficiency, \eff{\trk}, is also evaluated for muons using
a tag-and-probe method on the $\z \to \dimu$ data. The tag must
satisfy all the muon reconstruction and identification
requirements. The probe is reconstructed from a track segment in the
muon chambers that has been associated to a hit in the \ttt
sub-detector, which is not required in the track
reconstruction. Events with a tag and probe mass consistent with the
on-shell \z mass are used. The tracking efficiency is evaluated as the
number of events with a reconstructed probe track over the total
number of events. For lower \pt tracks, masses consistent with the
\jpsi are used.

The $\jpsi \to \dimu$ topology differs from the $\z \to \ditau$
topology in both pseudorapidity and momentum, and so the \jpsi muon
tracking efficiencies are evaluated in bins of both variables. The
muon tracking efficiency is found to vary between $85\%$ and
$93\%$. The uncertainty on the tracking efficiency is given by the
statistical precision and the knowledge of the purity of the sample of
$\jpsi \to \dimu$ candidates. The purity is estimated by fitting the
di-muon invariant mass distribution of the $\jpsi \to \dimu$
candidates with a Crystal Ball function~\cite{crystal} to describe the
signal shape and a linear background. An alternative estimate is
obtained by fitting only the linear background on either side of the
di-muon resonance. The difference in the efficiency evaluated using
the two purity methods is taken as the systematic uncertainty.

All particles pass through approximately $20\%$ of a hadronic
interaction length of material prior to the final tracking station.
Early showering of hadrons reduces the hadron tracking efficiency
compared to the muon tracking efficiency.  An additional correction
factor to the muon tracking efficiency of \prc{84.3 \pm 1.5} for
hadrons is applied which has been estimated using the full detector
simulation, where the uncertainty on this correction corresponds to an
uncertainty of $10\%$ in the material budget~\cite{track}.

The electron tracking efficiency uses a tag-and-probe method on $\z
\to \die$ data events. The tag must satisfy all the electron
reconstruction and identification requirements and the probe is
selected as the highest energy \ecal cluster in the event not
associated with the tag. The purity of the sample is found, from
simulation, to depend on the \pt of the tag. The dependence of the
purity is fitted with signal and background templates obtained from
same-sign and opposite-sign events from data. No momentum information
is available for the probe, so the tag-and-probe technique only
provides an overall tracking efficiency for the electrons, which is
measured to be \prc{83 \pm 3}. The momentum dependence is taken from
$\z \to \die$ and $\z \to \ditau$ simulation. The electron tracking
efficiency uncertainty is taken from the fit uncertainty added in
quadrature to the statistical uncertainty.

The identification efficiency, \eff{\id}, is measured for muons with
the tag-and-probe method for the $\z \to \dimu$ data, using a
reconstructed track as the probe lepton and evaluated as a function of
the probe momentum. For low momenta the efficiency is evaluated using
a $\jpsi \to \dimu$ sample as a function of both probe pseudorapidity
and momentum. The muon identification efficiency is found to vary
between $93\%$ and $99\%$ in pseudorapidity and momentum. The muon
\eff{\id} uncertainty is evaluated with the same method used for the
muon \eff{\trk} uncertainty.

The electron identification efficiency is measured as a function of
probe momentum using the tag-and-probe method on $\z \to \die$ data
and simulation events. The isolation requirement introduces a bias of
$1\% - 4\%$ in data and reconstructed simulation and so simulation
without the isolation criteria is used instead. The electron
identification efficiency is found to vary between $85\%$ and $96\%$,
with an uncertainty in each bin estimated as the difference in the
biased efficiencies from data and simulation.

The hadron identification efficiency is determined using events
triggered on a single \velo track. The highest \pt track in each
minimum bias event is assumed to be a hadron, as verified by
simulation. The hadron identification efficiency is taken as the
fraction of tracks fulfilling the hadron identification
requirements. Although the minimum bias topology differs significantly
from the $\z \to \ditau$ topology, an efficiency dependence is
observed only in pseudorapidity and so the efficiency is evaluated as
a function of pseudorapidity and found to vary between $92\%$ and
$95\%$. The uncertainty for each bin of pseudorapidity is estimated as
the statistical uncertainty of the bin. A summary of the systematic
uncertainties is given in \tab{\ref{tab:unc}}.

%% file: Latex/results.tex
\begin{table}
  \caption{Systematic uncertainties expressed as a percentage
    of the cross-section for each $\z \to \ditau$ analysis
    stream. Contributions from acceptance \acc, branching fractions
    \br, number of background events $N_\bkg$, reconstruction
    efficiencies \eff{\rec}, and selection efficiencies \eff{\sel} are
    listed. The superscripts on $\eff{\trk}^{(i)}$ and $\eff{\id}^{(i)}$
    indicate the first or second \tl decay product candidate. The
    percentage uncertainties on the cross-section for
    $N_\bkg$ are quoted for each individual background, as well as the
    total background. The efficiency uncertainties are split in a similar
    fashion.\label{tab:unc}}
  \begin{center}
    \small
    \begin{tabular}{ll|rrrrr}
      \toprule
      \multicolumn{2}{c|}{\multirow{2}{*}{Stream}}
      & \multicolumn{5}{c}{$\Delta\sigma_{pp \to \z \to \ditau}~[\%]$} \\
      & & \mumu & \mue & \emu & \muh & \eh \\
      \midrule
      \multicolumn{2}{l|}{\acc}
      & $1.48$  & $1.61$ & $1.32$  & $1.10$ & $1.11$  \\
      \midrule
      \multicolumn{2}{l|}{\br}
      & $0.46$  & $0.32$ & $0.32$  & $0.32$ & $0.33$  \\
      \midrule
      \multicolumn{1}{l|}{\multirow{5}{*}{$N_\bkg$}}
      & $N_\qcd$
      & $4.33$  & $0.80$ & $3.08$  & $0.40$ & $0.92$  \\
      \multicolumn{1}{l|}{} & $N_\ewk$
      & $4.22$  & $1.54$ & $1.52$  & $0.40$ & $0.72$  \\
      \multicolumn{1}{l|}{} & $N_\ttbar$
      & $0.02$  & $0.08$ & $0.12$  & $0.00$ & $0.58$  \\
      \multicolumn{1}{l|}{} & $N_\ww$
      & $0.02$  & $0.14$ & $0.13$  & $0.09$ & $0.08$  \\
      \multicolumn{1}{l|}{} & $N_{\z}$
      & $8.00$  & $-$    &  $-$    & $0.22$ & $0.23$  \\
      \midrule
      \multicolumn{2}{l|}{Total $N_\bkg$}
      & $10.03$ & $1.75$ & $3.44$  & $0.61$ & $1.32$  \\
      \midrule
      \multicolumn{1}{l|}{\multirow{6}{*}{\eff{\rec}}}
      & \eff{\gec}
      & $0.10$  & $0.10$ & $0.10$  & $0.10$ & $0.10$  \\
      \multicolumn{1}{l|}{} & \eff{\trg}
      & $0.88$  & $0.71$ & $2.29$  & $0.72$ & $4.30$  \\
      \multicolumn{1}{l|}{} & $\eff{\trk}^{(1)}$
      & $0.71$  & $0.74$ & $3.67$  & $0.79$ & $3.67$  \\
      \multicolumn{1}{l|}{} & $\eff{\trk}^{(2)}$
      & $0.34$  & $3.67$ & $0.61$  & $1.76$ & $1.68$  \\
      \multicolumn{1}{l|}{} & $\eff{\id}^{(1)}$
      & $0.38$  & $0.28$ & $1.72$  & $0.29$ & $1.73$  \\
      \multicolumn{1}{l|}{} & $\eff{\id}^{(2)}$
      & $0.78$  & $0.18$ & $0.56$  & $0.03$ & $0.09$  \\
      \midrule
      \multicolumn{2}{l|}{Total \eff{\rec}}
      & $1.47$  & $4.21$ & $4.73$  & $2.08$ & $6.15$  \\
      \midrule
      \multicolumn{1}{l|}{\multirow{5}{*}{\eff{\sel}}} 
      & \eff{\kin}
      & $-$     & $1.04$ & $2.89$  & $-$    & $1.91$  \\
      \multicolumn{1}{l|}{} & \eff{\iso}
      & $1.79$  & $1.91$ & $3.19$  & $1.65$ & $2.75$  \\
      \multicolumn{1}{l|}{} & \eff{\dphi}
      & $1.08$  & $1.03$ & $1.86$  & $0.60$ & $0.97$  \\
      \multicolumn{1}{l|}{} & \eff{\ips}
      & $2.70$  & $-$    &  $-$    & $1.92$ & $2.85$  \\
      \multicolumn{1}{l|}{} & \eff{\apt}
      & $2.03$  & $-$    &  $-$    & $-$    & $-$     \\
      \midrule
      \multicolumn{2}{l|}{Total \eff{\sel}}
      & $3.97$  & $2.41$ & $4.69$  & $2.60$ & $4.50$  \\
      \midrule
      \multicolumn{2}{l|}{Total systematic}
      & $11.13$ & $5.41$ & $7.56$  & $3.88$ & $7.88$  \\
      \bottomrule
    \end{tabular}
  \end{center}
  \vspace{5cm}
\end{table}

\section{Results}\label{sec:res}

The cross-sections for each analysis stream are determined using
\equ{\ref{equ:xs}}, the values given in \tab{\ref{tab:mea}}, and the
systematic uncertainties presented in \tab{\ref{tab:unc}}. The results
are
\begin{equation*}
  \begin{aligned}
    \sigma_{pp \to \z \to \ditau} ~ (\mumu) ~ &= 77.4 \pm           10.4 \pm 8.6 \pm 2.7 ~\pb \\
    \sigma_{pp \to \z \to \ditau} ~ (\mue ) ~ &= 75.2 \pm \phantom{1}5.4 \pm 4.1 \pm 2.6 ~\pb \\
    \sigma_{pp \to \z \to \ditau} ~ (\emu ) ~ &= 64.2 \pm \phantom{1}8.2 \pm 4.9 \pm 2.2 ~\pb \\
    \sigma_{pp \to \z \to \ditau} ~ (\muh ) ~ &= 68.3 \pm \phantom{1}7.0 \pm 2.6 \pm 2.4 ~\pb \\
    \sigma_{pp \to \z \to \ditau} ~ (\eh  ) ~ &= 77.9 \pm           12.2 \pm 6.1 \pm 2.7 ~\pb \\
  \end{aligned}
\end{equation*}
where the first uncertainty is statistical, the second uncertainty is
systematic, and the third is due to the uncertainty on the integrated
luminosity.

A global fit is performed using a best linear unbiased
estimator~\cite{blue} including correlations between the final states,
and a combined result of
\begin{equation*}
  \sigma_{pp \to \z \to \ditau} = 71.4 \pm 3.5 \pm 2.8 \pm 2.5 ~\pb
\end{equation*}
is obtained, with a $\chi^2$ per degree of freedom of $0.43$. The
statistical uncertainties are assumed to be uncorrelated as each
analysis stream contains mutually exclusive datasets. The luminosity
and any shared selection or reconstruction efficiencies are assumed to
be fully correlated.

\begin{figure}[h]
  \begin{center}
    \includesvg[width=\columnwidth]{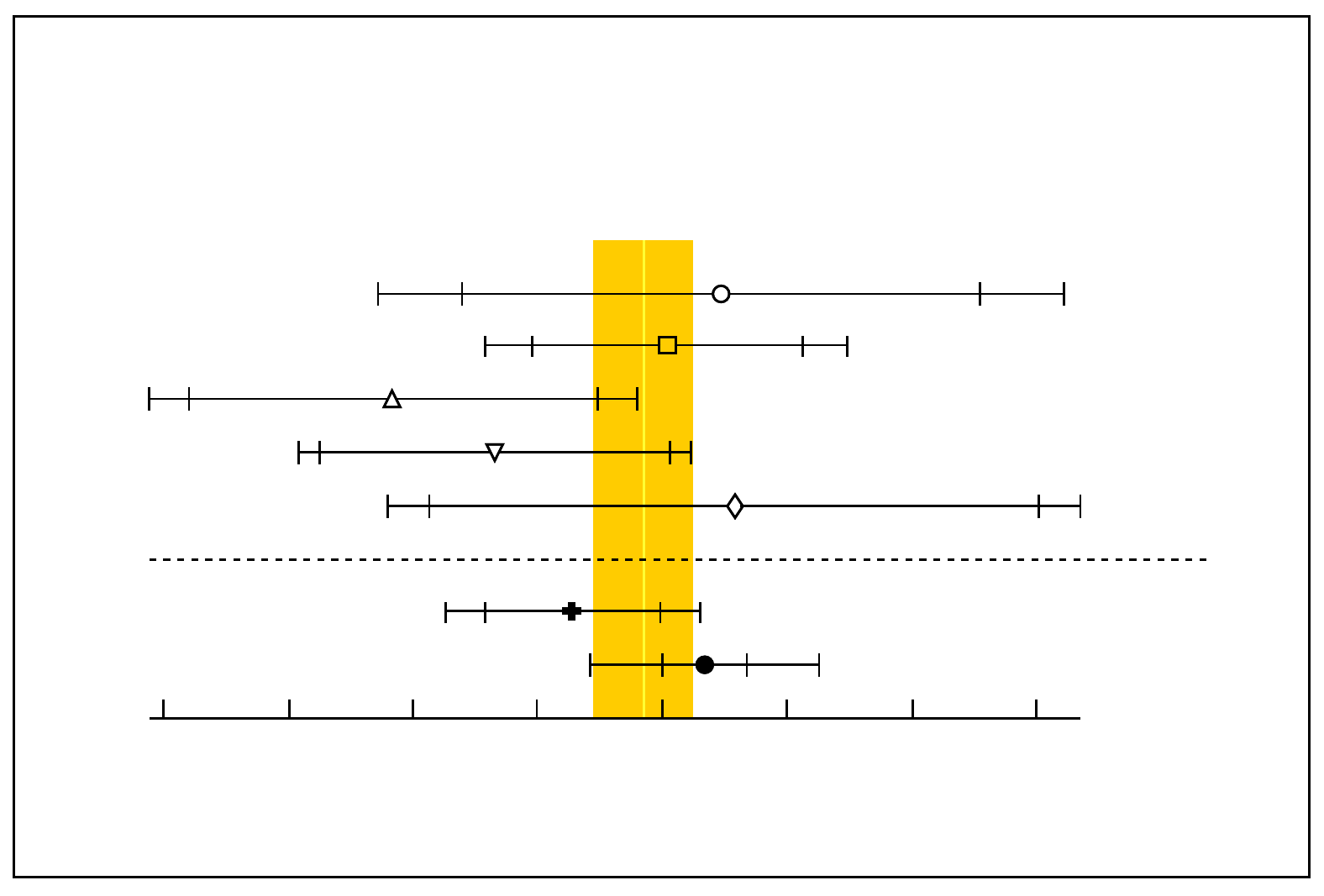}
  \end{center}
  \caption{Measured cross-sections for the \z decaying to the final
    states \mumu, \mue, \emu, \muh, and \eh (open points) compared
    with theory (yellow band) and the combined $\z \to \ditau$ and
    \lhcb $\z \to \dimu$ measurements (closed points) where \pt and
    $\eta$ are the transverse momentum and pseudorapidity of the
    leptons, and $M$ is the di-lepton invariant mass. The inner error
    bars represent statistical uncertainty while the outer error bars
    represent combined statistical, systematic, and luminosity
    uncertainties. The central theory value is given by the light
    yellow line while the associated uncertainty by the orange
    band.\label{fig:sum}}
\end{figure}

A graphical summary of the individual final state measurements, the
combined measurement, the $\z \to \dimu$ measurement of
\rfr{\cite{lhcb}}, and a theory prediction is shown in
\fig{\ref{fig:sum}}. The theory calculation uses \dynnlo~\cite{dynnlo}
with the \mstw next-to-next-leading-order (NNLO) PDF set~\cite{mstw},
and is found to be $74.3^{+1.9}_{-2.1}~\pb$.

The ratio of the combined cross-section to the \lhcb $\z \to \dimu$
cross-section measurement~\cite{lhcb} is found to be
\begin{equation*}
  \frac{\sigma_{pp \to \z \to \ditau}}{\sigma_{pp \to \z \to \dimu}}
  = 0.93 \pm 0.09
\end{equation*}
where the uncertainty is the combined statistical, systematic, and
luminosity uncertainties from both measured cross-sections, which are
assumed to be uncorrelated.

\afterpage{\clearpage}

%% file: Latex/conclusion.tex
\section{Conclusions}\label{sec:con}

Measurements of inclusive $\z \to \ditau$ production in $pp$
collisions at $\sqrt{s} = 7~\tev$ in final states containing two
muons, a muon and an electron, a muon and a hadron, and an electron
and a hadron have been performed using a dataset corresponding to an
integrated luminosity of $1028 \pm 36~\ipb.$ The cross-sections for
the individual states have been measured in the forward region of $2.0
\leq \eta^\tau \leq 4.5$ with $\pt^\tau > 20~\gevc$ and $60 <
M_{\tau\tau} < 120~\gevcc$, and a combined result calculated. The
results have been compared to Standard Model NNLO theory predictions
and with the \lhcb $\z \to \dimu$ cross-section measurement. The
individual measurements, the combined result, the $\z \to \dimu$
cross-section, and the theory prediction are all in good
agreement. The ratio of the $\z \to \dimu$ cross-section to the $\z
\to \ditau$ cross-section is found to be consistent with lepton
universality.

%% file: Latex/acknowledgements.tex
\section*{Acknowledgements}

\noindent We express our gratitude to our colleagues in the CERN
accelerator departments for the excellent performance of the LHC. We
thank the technical and administrative staff at the LHCb
institutes. We acknowledge support from CERN and from the national
agencies: CAPES, CNPq, FAPERJ and FINEP (Brazil); NSFC (China);
CNRS/IN2P3 and Region Auvergne (France); BMBF, DFG, HGF and MPG
(Germany); SFI (Ireland); INFN (Italy); FOM and NWO (The Netherlands);
SCSR (Poland); ANCS/IFA (Romania); MinES, Rosatom, RFBR and NRC
``Kurchatov Institute'' (Russia); MinECo, XuntaGal and GENCAT (Spain);
SNSF and SER (Switzerland); NAS Ukraine (Ukraine); STFC (United
Kingdom); NSF (USA). We also acknowledge the support received from the
ERC under FP7. The Tier1 computing centres are supported by IN2P3
(France), KIT and BMBF (Germany), INFN (Italy), NWO and SURF (The
Netherlands), PIC (Spain), GridPP (United Kingdom). We are thankful
for the computing resources put at our disposal by Yandex LLC
(Russia), as well as to the communities behind the multiple open
source software packages that we depend on.

%% file: Latex/bibliography.tex
\newboolean{articletitles}
\setboolean{articletitles}{true}
\bibliographystyle{lhcb}
\bibliography{Latex/bibliography}